\documentstyle[12pt]{article}
\newcommand{\be}{\begin{equation}}
\newcommand{\ee}{\end{equation}}
\begin{document}
\title{Spin Diffusion in 2D XY Ferromagnet with Dipolar
Interaction}
\author{A.Kashuba$^{ab}$, Ar.Abanov$^b$ and V.L.Pokrovsky$^{ab}$}
\maketitle
{\it $^a$Landau Institute for Theoretical Physics,
Kosygin 2, Moscow 119740, Russia \\
$^b$Department of Physics, Texas A\&M University,
College Station, Texas 77843-4242, USA}

\begin{abstract}
In the ordered phase of 2D XY ferromagnet the dipole force 
induce a strong interaction between spin-waves in the long 
wavelength limit. This interaction leads to the transformation
of the spin-wave excitation into a new soft-mode excitation 
in an intermediate range of wavelengths limited in magnitude 
and direction; and into an anomalous anisotropic diffusion 
mode excitation at long wavelengths. The dissipation of a 
spin-wave at short wavelengths is found to be highly 
anisotropic.
\end{abstract}


\newpage 

The conventional condensed matter theory deals with
elementary excitations and their interactions. The
excitations such as electrons, phonons, spin-waves etc.
have a propagating, wave-like nature. The momentum
$\bf p$ and the energy $\omega$ of a single excitation
are related by the dispersion relation (spectrum):
$\omega=\epsilon({\bf p})$. A weak interaction changes
slightly the spectrum and leads to a finite
life-time of the excitations \cite{AGD}. Effect of a
strong interaction is not so universal. Migdal
\cite{migdal} has shown that a strong electron-phonon
interaction renormalizes substantially electron velocities
and, if it exceeds a critical value, destroys the electron
quasi-excitations.
Recently a strong interaction of electrons in 2 dimensions
with gauge fields has been shown to result in non-Fermi-liquid
behavior \cite{correlectron}.

In the long wavelength (hydrodynamic) limit quasi-excitations
turn into classical modes, such as sound or spin-wave. Not
only the propagating waves, but also particle, heat and
spin diffusion can be considered as hydrodynamic modes.
The interaction between these modes has been shown to be
substantial in the critical region \cite{kadanoff};
in the dynamics of liquid crystals \cite{kats}, \cite{zeyher};
and in the dynamics of a Charge Density Wave interacting
with impurities \cite{gruner}. In all these systems the
interaction leads to a drastic reconstruction of the
dispersion relation.

In this Letter we solve the experimentally feasible magnetic
model, in which a strong interaction between spin-waves leads
to the replacement of the propagating spin-wave by a
diffusion mode and to appearance of a new soft-mode in
a range of momentum. This model is the two-dimensional
XY ferromagnet with the dipolar interactions between spins.

A spin diffusion mode appears naturally in the paramagnetic 
phase and in the vicinity of the Curie point \cite{Schwabl}. 
We consider a low temperature ordered
phase, where no diffusion is expected but rather a propagating 
and weakly dissipating spin-wave mode. In a 3D ferromagnet 
the exchange interaction between
spin-waves vanishes in the long wavelength limit \cite{dyson}.
The dipole force generates three spin-wave processes
\cite{Schwabl,kag-chub} and violates the total spin conservation
law. This interaction is dominant in the dissipation of a
spin-wave via decaying it into the two spin-waves or via
merging it with other spin-wave. But dissipation is weak.

In a 2D XY ferromagnet at low temperatures the dipolar
interaction is relevant in the long wavelength limit,
even despite of the low density of spin-waves. It was
shown by one of the authors \cite{kashuba} that dipolar
force induces an anomalous anisotropic scaling of
spin-spin correlations in the ordered phase.
In this article we find an analogous dynamical scaling.

The dipolar force stabilizes the ferromagnetic long-range
order \cite{MPF}, suppressing strong XY thermal fluctuation.
Therefore, we represent the unit vector field of magnetization
${\bf S}$ by the two spin-wave fields -- in-plane
$\phi({\bf x},t)$ and out-of-plane $\pi({\bf x},t)=S^z$:
\be {\bf S}=\left(-\sqrt{1-{\cal \pi}^2}\sin{\phi};
\sqrt{1-{\cal \pi}^{2}}\cos{\phi};\pi\right) \label{S}
\ee
where both $\pi$ and $\phi$ are small. The field
$\pi(t,{\bf x})$ is canonically conjugated to the field
$\phi(t,{\bf x})$.

The Hamiltonian of a 2D XY ferromagnet contains three
terms: the exchange and the anisotropy energies:
$$H_{EA}=\sum_{\omega\,{\bf p}}(J{\bf p}^2
|\phi_{\omega{\bf p}}|^2+\lambda
|\pi_{\omega{\bf p}}|^2 )/2,$$
as well as the dipolar force term 
(see e.g. \cite{kashuba}):
\be H=H_{EA}+g\sum_{\omega\,{\bf p}}\left| p_x
\phi_{\omega{\bf p}}+p_y\left(\phi^2/2\right)_{
\omega{\bf p}}\right|^2/2|{\bf p}|,
\label{Ham}
\ee
with the corresponding couplings being -- the exchange
constant $J$, the anisotropy $\lambda$ and the dipole
constant $g$ \cite{g};
The Hamiltonian
(\ref{Ham}) is written in terms of the Fourier transformed
fields -- $\phi_{\omega{\bf p}}$ and $\pi_{\omega{\bf p}}$,
the abbreviation - $\left(\phi^2/2\right)_{
{\bf p}\omega}$ - denotes the Fourier transform of
$\phi^2({\bf x},t)/2$. The anisotropy is assumed to be weak:
$\lambda\ll J$. Therefore
the spin ${\bf S}$ averaged over scales larger than
$\sqrt{J/\lambda}$ turns into the plane.
The bare spin-wave spectrum extracted from
the quadratic part of the Hamiltonian (\ref{Ham}) reads:
\be
\epsilon^2({\bf p})=c^2\left(p^2+
p_0p_x^2/p\right),
\label{dis}\ee
where $p=|{\bf p}|$, the characteristic wave-vector $p_0=g/J$ 
and the spin-wave velocity $c=\sqrt{J\lambda}$. In the region 
$p_0\ll p\ll\sqrt{\lambda/J}$ spin-waves have the linear, 
acoustic-like spectrum. We call this spherical shell of 
momentum the $\cal A$-shell.

The non-linear part of the Hamiltonian
(\ref{Ham}) $H_{int}$
contains the three-leg and four-leg vertices.
In particular the three-leg vertex is:
\be
f({\bf p}_1, {\bf p}_2, {\bf p}_3)=\sum_{i=1}^3
p_{ix}p_{iy}/\mid {\bf p}_i\mid
\label{vertex}
\ee
Since the sum of momenta entering the vertex is zero it depends
on two momenta. Further we denote it as $f({\bf p}_1, {\bf p}_2)$.

The dynamical equation associated with the Hamiltonian
(\ref{Ham}) reads:
\be
\left({\omega^2\over\lambda}-{\epsilon^2({\bf p})\over
\lambda}+i{\omega\over\Gamma_0}\right)
\phi_{\omega{\bf p}}=\frac{\delta H_{int}}{\delta
\phi_{-\omega-{\bf p}}}+\eta_{\omega{\bf p}}. \label{Eqn}
\ee
The thermal noise $\eta_{\omega{\bf p}}$ and the bare
dissipation coefficient $\Gamma_0^{-1}$ generate the
stochastic dynamics in (\ref{Eqn}). The noise-noise
corelation function is related to the bare dissipation
coefficient: $\langle|\eta_{\omega{\bf p}}|^2\rangle=
2T/\Gamma_0$. It vanishes at $T=0$.
We do not consider external sources of dissipation,
spin-waves dissipate due to interaction generated by dipole
force. Hence, we set
$\Gamma_0^{-1}=+0$, and look for the generated
dissipation: $\Gamma^{-1}(\omega,{\bf p})$.

The Green function $G(\omega,{\bf p})$ is defined as the
linear response of the magnet to an external magnetic
field with the same frequency and wave-vector. The l.-h.-side
of Eq.\ref{Eqn} represents the
inverse of the bare Green function $\lambda^{-1}G_0^{-1}
(\omega,{\bf p})$.
According to the Fluctuation-Dissipation Theorem, the
imaginary part of the Green function multiplied by the
factor $T/\omega$ is the spin-spin correlation function
$D(\omega,{\bf p})$.
For convenience we use the dissipation function defined as:
$b(\omega,{\bf p})=\lambda/\Gamma(\omega,{\bf p})$. The
self-energy term $\Sigma(\omega,{\bf p})$ equals
$G^{-1}_0(\omega,{\bf p})-G^{-1}(\omega,{\bf p})$ by
definition. We notify the real and the imaginary part
of the self-energy term as: $\Sigma =a^2(\omega,{\bf p})
-i\omega b(\omega,{\bf p})$. Thus, the Green function
reads:
\be G^{-1}(\omega,{\bf p})=\omega^2-\epsilon^2
({\bf p})-a^2(\omega,{\bf p})+i\omega b(\omega,{\bf p}),
\ee
whereas the spin-spin correlation function reads:
\be D(\omega,{\bf p})=\frac{b(\omega,{\bf p})}{\left[
\omega^2-\epsilon^2({\bf p})-a^2(\omega,{\bf p})
\right]^2+\omega^2b^2 (\omega,{\bf p})} \ee
(we refer the factor $T$ to vertices).
We use the reduced temperature $t=T/4\pi J$
as a small and $L=\log(\sqrt{J\lambda}/g)$ as a large 
parameters (the latter means large ${\cal A}$-shell).

We employ the standard, so-called Janssen-De-Dominicis
Functional method \cite{martin} to account for the
non-linear terms in the stochastic Langevin equation
(\ref{Eqn}). This method generates a diagrammatic
expansion in powers of the bare vertices, associated 
with $H_{int}$. The main contribution to the
self-energy is given by the one-loop diagram shown 
in Fig.1a, 1b. Our theory is valid only if the
temperature is small: $$t\log(\sqrt{J\lambda}/g)=tL\ll 1.$$
Under this condition the two-loop corrections are small and
the diagram Fig.1b contributes to a negligible change
of the spectrum (\ref{dis}) \cite{zinn}. Such neglecting 
of the two-loop diagrams (vertex correction) was a major
assumption in the so-called mode-coupling methods 
\cite{Schwabl}. Later we proove this assumption.
Thus, the Dyson equation for our problem is as follows:
\begin{eqnarray}
\Sigma(\Omega,{\bf q})=2g^2\lambda^3
T\int{d^2{\bf p}\over (2\pi)^2}\int{d\omega\over 2\pi}
f^2({\bf p},{\bf q})\nonumber\nopagebreak\\
D(\omega,{\bf p})G(\omega+\Omega,{\bf p+q}),
\label{rea}
\end{eqnarray}
The functions
$b(\omega,{\bf p})$ and $a(\omega,{\bf p})$ are even
in both arguments \cite{AGD}.
The imaginary part of the self-energy 
is odd in frequency $\Omega$. Hence,
the equation for the dissipation function reads:
\begin{eqnarray}
b(\Omega,{\bf q})=g^2\lambda^3T
\int{d^2{\bf p}\over (2\pi)^2}\int{d\omega\over 2\pi}
f^2({\bf p},{\bf q})\nonumber\nopagebreak\\
D(\omega,{\bf p})D(\omega+\Omega,{\bf p+q}).
\label{ME}
\end{eqnarray}

The integrand in (\ref{ME}) is positive. Thus, the main
contribution to $b(\Omega,{\bf q})$ comes from the region,
where poles of the two $D$-functions coincide. The
function $D(\omega,{\bf p})$ has poles at:
$\omega\approx\pm\epsilon({\bf p})$, in the
${\cal A}$-shell \cite{a}. Following the terminology
of the field theory we call the surface
$\omega^2=\epsilon^2({\bf p})$ the mass-shell.
The dissipation in the ${\cal A}$-shell is small and
the $D$-function can be represented as a sum of 
$\delta$-functions:
\be 
D(\omega,{\bf p})\approx\sum_\pm
{\pi\over 2\epsilon^2({\bf p})}\delta(\Delta\omega_\pm),
\label{D1} 
\ee
where $\Delta\omega_\pm=\omega\pm\epsilon({\bf p})$
measures the deviation from the mass-shell.
After integrating $\omega$ out from the Eq.\ref{ME}
with the $D$-functions from (\ref{D1}) we recover the
Fermi Golden Rule for the probability of the spin-wave
decay and scattering processes.

Looking for the long wavelength quasi-excitations we need 
the self-energy at very small momenta $q\ll p_0$, which we 
denote as $\Sigma_0$. We expect quasi-excitations to be soft:
$\Omega\ll cq$, and we restrict the wave-vector ${\bf q}$ of 
quasi-excitation to be directed along the magnetization: 
$|q_x|\ll q$.
The essential contribution to the integral in Eq.\ref{rea}
comes from the internal momentum $p$ being in the $\cal A$-shell
and the internal frequency $\omega=\epsilon({\bf p})$. 
Taking the integral over $\omega$, with the $D$-function 
from Eq.(\ref{D1}), we find \cite{a}:
\begin{equation}
\Sigma_0={c^2p_0^2t\over 4\pi}\int
{c^4p^3dp\over\epsilon^4({\bf p})}{\Omega\sin^2(2\psi)
 d\psi\over \Omega-cq\cos\psi+ib_1}, \label{B0}
\end{equation}
where $b_1$ is the dissipation function $b(\omega,{\bf p})$ 
of a spin-wave inside the ${\cal A}$-shell and $\psi$ is the
direction of the internal spin-wave: $\sin\psi=p_x/p$. Note, 
that we added a $\Omega$-independent contribution from
the diagram Fig.1.b to the self-energy \ref{B0}.
$\mbox{Re}\Sigma_0$ vanishes in the static limit
$\Omega =0$.

If $cq\gg b_1$  we make the integral over $\psi$ in 
Eq.\ref{B0} to find:
\be  \Sigma_0(\chi)=c^2p_0^2tL
\cos^2\chi\exp({-2i\chi}),\label{B00} \ee
where $\chi$, defined by the equation $\cos\chi=\Omega/cq$,
measures the deviation from the mass shell.
If $q$ is so small that $cq\ll b_1$ the Eq.\ref{B0} implies 
the $q$-independent dissipation constant:
\be b_0=c^2p_0^2tL\int{d\psi\over 4\pi}
{\sin^2(2\psi)\over b_1(\psi)}. \label{MR} \ee
In this calculation we have used the fact that the
dissipation of a spin-wave in the ${\cal A}$-shell $b_1$ 
depends only on the angle $\psi$ between the direction 
of magnetization and the spin-wave wave-vector ${\bf p}$, 
which we prove below.

Anyway, we need $b_1(\psi)$. An unusual feature of our
theory is that the dissipation process in the
${\cal A}$-shell is mediated by an off-mass-shell,
virtual spin-wave. Indeed, the Eq.\ref{dis}
does not allow for decay or merging processes.
Alternatively, the dissipation of a
spin-wave in the ${\cal A}$-shell propagating in the
direction specified with the angle $\psi$ 
($\sin\psi=q_x/q$) is mediated by an internal virtual 
spin-wave in (9), with a momentum $p\ll p_0\ll q$ 
and a frequency $\omega <cp$. The integration over 
$\omega$ with one of 
the $D$-functions in (\ref{ME}) taken in the form 
(\ref{D1}), leads to a following equation:
\be b_1={c^2p_0^2tf^2({\bf 0},{\bf q})\over
8\pi q^2}\int d^2{\bf p}D(\epsilon({\bf p+q})-
\epsilon({\bf q}),{\bf p}) \label{sss} \ee
Since $\omega=\epsilon({\bf p+q})-\epsilon({\bf q})$
we conclude that $\omega=cp\cos\Phi$, where
$\Phi=\psi-\varphi$ is the angle between the vectors
${\bf q}$ and ${\bf p}$. Invoking the definition
of the angle $\chi$ for virtual
spin-wave, we find that $\chi=\Phi$. Now, we look onto
Eq.\ref{sss} in more details:
\be 
b_1(\psi)={c^2p_0^2t\over 2\pi}\mbox{Im}\int
{\sin^2\psi\cos\psi\ dpd\varphi\over p^2\sin^2\psi+
p_0p\varphi^2+\Sigma_0(\psi)/c^2}. 
\label{B1}
\ee
Note, that $\varphi^2\sim p/p_0\ll 1$ in (\ref{B1}).
Thus, $\chi=\psi$. In other words, the
dissipation of a short wavelength spin-wave propagating
in the direction $\psi$ is determined by the scattering
on the long wavelength virtual spin-wave, which lies
on the specific distance off the mass-shell: $\omega/cp=
\cos\psi$. The integration over $p$ in (\ref{B1})
is confined towards the crossover region: 
$p\sim p_c=p_0\sqrt{tL}$.

Substituting $\Sigma_0(\psi)$ from Eq.\ref{B00} into 
Eq.\ref{B1} we find  the anisotropic dissipation of a 
spin-wave mode in the ${\cal A}$-shell:
\be b_1(\psi)=\beta_1 t^{3/4}cp_0{\sin^{3/2}(2\psi)
\sin(\psi/2)\over L^{1/4}\cos\psi},
\label{B11}\ee
where the direction of the spin-wave is limited to
the fundamental quadrant: $0<\psi<\pi/2$. We found:
$\beta_1=\Gamma^2(1/4)/4\sqrt{2\pi}\approx 1.31$.

Let us return to very low momenta 
$p\ll b_1/c$. Plugging  Eq.\ref{B11} into Eq.\ref{MR} 
one finds:
$ b_0=\beta_0 cp_0t^{1/4}L^{5/4}$, where $\beta_0
\approx 1.24$ was found numerically.
The condition $cp_{DM}\sim b_1$ defines the
crossover wave-vector: $p_{DM}\sim p_0t^{3/4}/L^{1/4}$,
between the self-energies Eq.\ref{B00} and Eq.\ref{MR}.

The dissipation functions (\ref{B00},\ref{MR} and \ref{B11}) 
conclude the self-consistent solution of the Dyson equation 
(\ref{rea}, \ref{ME}).

We verify that the two-loop correction (see Fig.1c)
is negligible. Note, that the
main contribution to the diagram Fig.1c comes, if the
two internal momenta ${\bf p}_1$ and ${\bf p}_2$ are
restricted to the ${\cal A}$-shell. Inside the
${\cal A}$-shell the Green and the D-functions
live on the mass-shell. But, the prohibition of the
three spin-wave processes leaves the Green function
$G(\omega_1-\omega_2,{\bf p_1-p_2})$ to be off the
mass-shell. The Green function
off the mass-shell is small in temperature. A simple
counting shows, then, that the two-loop dissipation
function is $b_0'=b_0(1+t^{1/4}O(t))$. The function
$b_1'(\psi,q)-b_1(\psi)$, which represents the
two-loop corrections for $b_1$, is small in $t^{1/4}$,
and is also small in the ratio $p_0/q$.

Having the explicit expression for the self-energy
we can analyse the dispersion relation --
$\omega^2=\epsilon^2({\bf p})+\Sigma(\omega, {\bf p})$
-- in the range
of small $\omega$ and $p$. New results are expected
for the region $p<p_c=p_0\sqrt{tL}$ in which $\Sigma_0$
becomes comparable with $\epsilon^2({\bf p})$.
In a range of momentum $p_{DM}\ll p\ll p_c$
and angles $\psi\ll\sqrt{p_0tL/p}$ we find a new
propagating soft mode with the dispersion:
\be
\omega = cp(p^2 + p_0p\psi^2)^{1/2}/p_0\sqrt{tL}
\label{soft}
\ee
The dissipation of the soft mode grows to the boundary
of the region and becomes of the order of its energy
at $\psi\sim \sqrt{p_0tL/p}$ or $p\sim p_{DM}$. There is
no soft mode beyond the indicated range. The spin-wave 
mode persists at $p>p_0\sqrt{tL}$. In a range $p\ll p_{DM}$ and 
small angles a new diffusion mode occurs with the 
dispersion:
\be
\omega = -i\epsilon^2({\bf p})t^{-1/4}L^{5/4}/\beta_0cp_0
\label{diff}
\ee
The angular range of the diffusion mode increases with
decreasing $p$ and captures the entire circle at
$p<p_0tL$.

At even smaller wavelengths $p<p_{A}\ll p_{DM}$ the interaction 
between diffusion modes should be taken into account (
$p_{A}$ is the anomalous diffusion set up wave-vector 
\cite{AKP}).
The problem can be solved by the renormalization group method
\cite{AKP}. The growing interaction, although leaving invariant
the diffusive nature of the spin propagation, changes the
dispersion. For the propagation along the spontaneous
magnetization it is: $\omega\propto -ip^{47/27}$, whereas
for the propagation in the perpendicular direction it is:
$\omega\propto -ip^{47/36}$. This is a dynamic analog
of the non-Gaussian fixed point found in \cite{kashuba}.

In conclusion, we discuss the experimental feasibility
of the above considered effects in the epitaxial magnetic films. 
The main difficulty is
that all of them are confined to rather long wavelengths. 
Therefore even weak in-plane anisotropy can suppress them. 
One need to use substrates with the six-fold symmetry axis. 
The six-fold anisotropy is much weaker than the tetragonal 
one. Moreover, it was predicted in \cite{PU} that
there exists a temperature interval in which the hexagonal
anisotropy vanishes at large distances. A proper substrate
is, for example, a (111) face of Cu, Au etc.
Unfortunately the epitaxial growth of ferromagnetic films
on this faces has not been realized so far. In a recent
work \cite{Rau} a growth of a
ultrathin Ru/C(1000) film has been reported. It
is the in-plane ferromagnet. We hope that further
sophistication of the experimental technics will allow
to observe the predicted dynamical behavior.

We are indebted to M.V. Feigelman, E.I. Kats and V.V. Lebedev
for useful discussions and indicating some references.

{\Large\bf Figure Caption}
\begin{itemize}
\item FIG.1. a) Main one-loop self-energy diagram; b) One-loop
four-leg vertex diagram; c) Two-loop self-energy diagram. 
Momenta of internal lines are indicated.
\end{itemize}

\begin{thebibliography}{22}
\bibitem{g} $g=2\pi(\gamma\mu_BS\sigma)^2$, where $\gamma$ is
the Lande-factor of magnetic ions, $S$ is its spin, $\mu_B$ is
the Bohr magneton and $\sigma$ is the area density of magnetic
ions in the film.
\bibitem{a} The $a$-function is negligible in the
${\cal A}$-shell. If one assumes, that
$a^2(\omega,{\bf p})\ll\epsilon^2({\bf p})$ inside the
${\cal A}$-shell, then from the Eq.\ref{rea} one finds:
$a^2(\epsilon({\bf q}),{\bf q})\sim tLc^2p_0\sqrt{p_0q}$,
justifying the assumption.
\end{thebibliography}
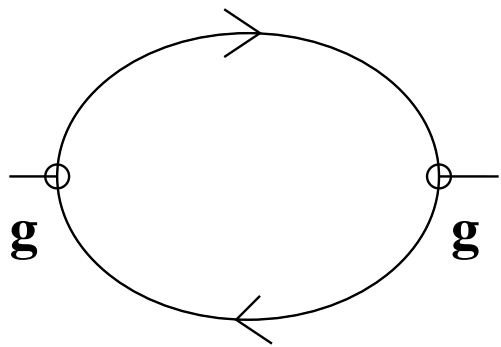
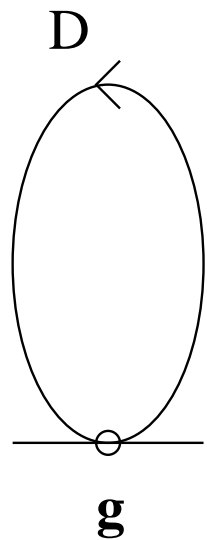
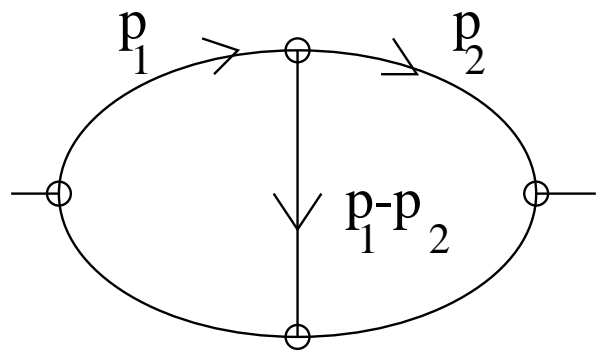
\end{document}